\documentstyle[prl,aps,multicol,epsf,rotate]{revtex}
\def\ffrac#1#2{\textstyle{#1\over#2}\displaystyle}
\begin{document}
\draft
%\preprint{XXXX}
%
\title{Exact Scaling Functions for Self-Avoiding Loops and 
Branched Polymers}
\author{John Cardy}
\address{University of Oxford, Department of Physics -- Theoretical
         Physics, 1 Keble Road, Oxford OX1 3NP, U.K. \\
         and All Souls College, Oxford.}
%
%\date{\today}
%
\maketitle
\begin{abstract}
It is shown that a recently conjectured form for the
critical scaling function for planar self-avoiding polygons weighted by their
perimeter and area
also follows from an exact renormalization group flow into the
branched polymer problem, combined with the dimensional reduction arguments of
Parisi and Sourlas. The result is generalized to 
higher-order multicritical points, yielding exact values for all
their critical exponents and exact forms for the associated scaling
functions. 

\end{abstract}
\begin{multicols}{2}
In two dimensions, exact results for critical exponents, which describe
power law dependence on a single relevant variable close to a
critical point, are commonplace.
By contrast, very few examples are known of exact
scaling functions which depend on
combinations of more than one such variable. In recent years
there has been considerable progress in problems involving counting
various restricted classes of random self-avoiding polygons\cite{tract}.
However, these are essentially
one-dimensional in nature, and so far no rigorous results exist for
the unrestricted case. Theoretically this latter case 
is perhaps more interesting because, in the scaling limit,
it corresponds to an isotropic field theory, the
$n\!\to\!0$ limit of the O$(n)$ model. While much exact information is known
about such critical theories in two dimensions, up to now no exact,
nontrivial,
scaling functions of more than one intensive thermodynamic variable
(such as the equation of state) have been found.

Recently Richard, Guttmann and Jensen\cite{RGJ} (hereinafter referred 
to as RGJ) have conjectured 
the exact form of such a scaling function for unrestricted
self-avoiding polygons in the plane. 
In the ensemble in which each link of the polygon is weighted with
fugacity $x$, this problem exhibits a critical point at some
value $x=x_c$. As $x\to x_c$ from below, the mean perimeter 
$\langle N\rangle$, mean
square radius of gyration $\langle R^2\rangle$ and mean area $\langle
A\rangle$ all diverge. 
Such self-avoiding loops provide a simple model
for two-dimensional vesicles\cite{LSF}, and in that context it is natural to
weight the ensemble according to the area of each loop, thus defining
the generating function for rooted loops
\begin{displaymath}
G^{(r)}(x,g)=\sum_{N,A}p^{(r)}_{N,A}x^N{\rm e}^{-gA}
\end{displaymath}
where $p^{(r)}_{N,A}$ is the number of such loops of a given perimeter and
area which pass through a given link of the lattice, 
and $g$ ($\equiv-\ln q$ in the notation of
RGJ) is the pressure difference
across the vesicle wall, in units of $kT$. From the point of view of
critical phenomena, this ensemble has a multicritical point at $x=x_c$,
$g=0$, and in its neighborhood one expects\cite{LSF} that the singular part of
$G$ has the scaling form (in the notation of RGJ)
\begin{equation}
\label{scaling}
G_{\rm sing}^{(r)}(x,g)=g^\theta\,F\left((x_c-x)g^{-\phi}\right)
\end{equation}
where $\theta$ and $\phi$ are related to the conventional exponents 
$\nu$ and $\alpha$ by $\theta/\phi=1-\alpha$ and $\phi=1/2\nu$,
based on the assumptions that, at $g=0$,
the singular part of the
mean number of such rooted loops behaves like $(x_c-x)^{1-\alpha}$
and the mean area $\langle A\rangle \sim\langle
R^2\rangle\sim (x_c-x)^{-2\nu}\sim\langle N\rangle^{2\nu}$. In addition
$\alpha$ is related to $\nu$ by hyperscaling: $\alpha=2-2\nu$. There is ample
evidence from enumeration and other methods\cite{num} to support these
assumptions and the theoretical value\cite{nien} $\nu=\ffrac34$, so that
$\theta=\ffrac13$ and $\phi=\ffrac23$. 

RGJ, in analogy with similar but analytically tractable enumeration
problems\cite{tract}, 
assume that, as function of $x$ and $q={\rm e}^{-g}$, $G^{(r)}$
satisfies some $q$-algebraic functional equation of finite
degree. Together with the assumed values for $\theta$ and $\phi$, this
leads, in the limit $q\to1$, to a Riccati equation 
for the scaling function $F(s)$, whose solution is
\begin{equation}
\label{airy}
F(s)=b_0{d\over ds}\ln\,{\rm Ai}\left(b_1s\right)
\end{equation}
where ${\rm Ai}(x)\propto \int_{-\infty}^\infty {\rm e}^{{\rm i}xt
+{\rm i}t^3/3}dt$ is the Airy function, and $b_0$, $b_1$ are
non-universal constants. (\ref{airy}) determines exactly, for example, the
universal moment ratios $\langle A^p\rangle/\langle A\rangle^p$ as
$x\to x_c$, and
RGJ produce convincing evidence, based on extensive
enumerations, that these predictions are indeed correct. 

In this Letter, it is pointed out that (\ref{airy}) also follows from 
a completely different argument, which invokes the
physical reasoning of Ref.~\cite{LSF} 
to relate this problem to that of branched
polymers, combined with the dimensional reduction arguments of Parisi
and Sourlas\cite{PS} which map this latter problem to that of the Yang-Lee
edge singularity in two fewer dimensions. From this point of view, the
Airy integral then arises as the scaling limit of the
partition function of the Yang-Lee problem in zero dimensions.
Moreover, in this approach, the values of the exponents
$\theta$ and $\phi$ emerge without any further assumptions. 

{}From this perspective it is simple to generalize the conjecture
of RGJ to higher-order multicritical points
of self-avoiding loops with $k$ relevant renormalization group (RG)
scaling variables $v_j$. These
may presumably be realized by tuning to critical values
many-body interactions between nearby portions
of the loop.
When this ensemble is, in addition, weighted by the area
of the loops, the generalization of 
(\ref{scaling}) to arbitrary $k$ is
\begin{equation}
\label{kscaling}
G^{(r)}_{\rm sing}=g^{\theta_k}\,F_k(v_1g^{-y_1/2},v_2g^{-y_2/2},\ldots)
\end{equation}
where $y_j$ is the 
RG eigenvalue of $v_j$. It will be argued that
the exact values for these, at the $k$th order multicritical point, are
\begin{equation}
\label{yjk}
y_j(k)=2(k-j+2)/(k+2)
\end{equation}
and that the exact form for $F_k$ is given in terms of
a generalized Airy integral 
$\int_C {\rm e}^{-V(\psi)/g}d\!\psi$ where 
$V(\psi)=\sum_{j=1}^kv_j\psi^j-\psi^{k+2}$.
The values given in (\ref{yjk}) agree with those derived from a
generalized Flory argument, applied to an ensemble in which the first
$k$ renormalized virial coefficients vanish. That they should be exact in two
dimensions was suggested earlier by Saleur\cite{saleur} on the basis of a
postulated $N\!=\!2$ supersymmetry. Here it is seen that
they follow from the mapping to a simple zero-dimensional problem.
However, our results, like those of Saleur, display a paradox in that
$k=2$, the obvious candidate for the $\Theta$-point, yields values for
the exponents which disagree with those of an exactly solvable model\cite{DS}
and with extensive numerical results. This is discussed in detail
later.

Finally, in the generalized Airy integral it is possible to 
take the limit $g\to0$, thus recovering
the scaling function in the original ensemble with no area-weighting.
This comes from the appropriate saddle-point of
$V(\psi)$, and therefore amounts to finding the root of a polynomial.
For example, for $k=2$ it is found that
$G^{(r)}_{\rm sing}=c_0v_2^{1/2}\,\Phi(c_1v_1/v_2^{3/2})$,
where $c_0$ and $c_1$ are non-universal constants, and the exact scaling
function, for $v_2>0$, is
\begin{equation}
\label{thetascaling}
\Phi^{>}(s)=\big(s+(s^2-1)^{1/2}\big)^{1/3}
+\big(s-(s^2-1)^{1/2}\big)^{1/3}
\end{equation}
where the branch cuts of the fractional powers are taken to
lie along the negative real axis. 

We now give more details of the reasoning leading to these results,
first discussing the case $k=1$ considered by RGJ.
The physical part of the argument is to regard the model with $g>0$ as
presenting a \em crossover \em phenomenon: the negative pressure causes the
vesicles to try to minimize their area, but there is competition between
this and the need to maximize the perimeter as $x\to x_c$.
Clearly for large $g$ at fixed $x<x_c$ the vesicles should collapse into
double-walled, branched structures, but assume, as suggested by the
numerical work of Ref.~\cite{LSF} that, at large enough
distance scales, this will also
happen as $x\to x_c$ for any fixed $g>0$, consistent with the
idea that there is an RG flow from the fixed
point describing self-avoiding loops to that corresponding to branched
structures. Thus (\ref{scaling}) has the form of a crossover scaling
function\cite{Fisher1}. We also assume that the structures 
which result are in the same universality class as conventional branched
polymers (lattice animals with no cycles), in which all trees
with the same total length are weighted equally. The theory of crossover
scaling\cite{Fisher1} then asserts that the scaling function $F(s)$ in
(\ref{scaling}) should have a singularity of the form 
$(s-s^*)^{1-\alpha_{BP}}$, where $\alpha_{BP}$ is the entropic
exponent for branched polymers. From this hypothesis various
interesting results follow, for example that as $g\to0$ the branched
polymer singularity should occur at $x=x_c(g)=x_c+s^*g^\phi+\cdots$, 
which has been confirmed in enumeration studies\cite{LSF},
as well as
various predictions for the $g$-dependence of the critical amplitudes.
In general, however, the functional form of
a crossover function is very difficult to calculate, since the
scaling variables at the new fixed point bear a complicated relationship to
the original ones, which requires following the RG
flow in detail. However in this example there are
considerable simplifications.

First state the problem in field-theoretic language, by writing the
area of a given loop as
\begin{equation}
\label{area}
A=\int\int G_{\lambda\sigma}(r_1-r_2)J_\lambda(r_1)
J_\sigma(r_2)d^2\!r_1d^2\!r_2
\end{equation}
where $J_\lambda$ is the density of a current of unit strength flowing
around the loop, and $G_{\lambda\sigma}$ is the Green function for a U$(1)$
gauge field ${\cal A}$. (\ref{area}) expresses the well-known fact
that, in a two-dimensional gauge theory, 
the expectation value of a Wilson loop 
obeys a strict area law. In Ref.~\cite{JCarea} it was used to
compute the mean area of self-avoiding loops. The generating
function $G^{(r)}$ for rooted loops is the derivative with respect to the
fugacity $x$ of 
\begin{displaymath}
Z=
\langle{\rm e}^{-gA}\rangle_{\rm SAL}=
\left\langle {\rm e}^{-\sqrt g\int J_\lambda {\cal A}_\lambda d^2\!r}
\right\rangle_{{\rm SAL},{\cal A}}
\end{displaymath}
where the average is taken over self-avoiding loops (SAL), 
each weighted by $x^N$,
and over the gauge field, with the usual weight 
$\exp(-\frac14\int F^{\lambda\sigma}F_{\lambda\sigma}d^2\!r)$. 
Self-avoiding loops may be mapped, in the standard way, to the
$n\to0$ limit of an O$(n)$ model. In this case it is useful to consider
complex O$(n)$ lattice spins ${\bf s}(r)$, so that the U$(1)$ current
$J_\lambda$ is the lattice version of $(1/2{\rm i})
({\bf s}^*\!\cdot\!\partial_\lambda{\bf s}-{\rm cc.})$, and the weights are
$\prod_{\rm nn}(1+x({\bf s}^*(r)\!\cdot\!{\bf s}(r')+{\rm cc.}))$.

The first observation is that, at $n=0$, there are no vacuum corrections
to the gauge field propagator $G_{\lambda\sigma}$ (as in the `quenched'
approximation in lattice gauge theories), so that, since ${\cal A}$
couples to a conserved current, the gauge coupling
$g$ is not renormalized. Its RG equation is simply
$dg/d\ell=2g$ to all orders, so that 
it flows to infinity, where the irrelevant variable $g^{-2}$ has
RG eigenvalue $-2$. The other simplification is that, in the limit where
$g$ is large, the total length of the branched polymer is one half
that of the perimeter of the loop, apart from corrections $o(N)$. Thus
the fugacity variable for the branched polymer problem,
close to $x_c$, is \em linearly \em related to the original fugacity $x$. 

Now recall the formulation of the branched polymer problem in $d$
dimensions, given by
Parisi and Sourlas\cite{PS}. This is the $n\to0$ limit\cite{Shapir}
of a theory of
fields $\psi_a$ ($a=1,\ldots,n$), weighted by ${\rm e}^{-S}$ where 
\begin{equation}
\label{replica}
S=\int\Big(\sum_a(\ffrac12(\nabla\psi_a)^2-\sum_{p\geq1}u_p\psi_a^p)
+v\sum_{ab}\psi_a^2\psi_b^2\Big)d^d\!r
\end{equation}
where $u_p$ is the fugacity for $p$ branches to meet at a given point,
and $v>0$ represents self-avoidance. After shifting the fields
to eliminate the $\psi_a^2$ term, rescaling, 
and retaining only the most relevant
couplings, the action takes the form
\begin{equation}
\label{rep}
S=\int\Big(\sum_a(\ffrac12(\nabla\psi_a)^2+V(\psi_a))+\Delta
\sum_{ab}\psi_a\psi_b\Big)d^d\!r
\end{equation}
where $V(\psi)=t\psi-\frac13\psi^3+O(\psi^4)$.
This theory is critical at some value $t\to t_c+$.
Parisi and Sourlas\cite{PS} argued that, at $n=0$, (\ref{rep}) is equivalent
to a supersymmetric theory. We follow the more direct transformation of
Ref.~\cite{JCsusy}: define new combinations of the fields
$\psi\equiv\frac12(\psi_1+(n-1)^{-1}\sum_2^n\psi_a)$,
$\omega\equiv\Delta^{-1}(\psi_1-(n-1)^{-1}\sum_2^n\psi_a)$,
together with $n-2$ other fields $\chi_a$ ($a=3,\ldots,n$) which are
linear combinations of $(\psi_2,\ldots,\psi_n)$ orthogonal to 
$\sum_2^n\psi_a$. Discarding terms higher than quadratic order in $\omega$
and the $\chi_a$ (which may be shown to be irrelevant), the action
has the form, at $n=0$,
\begin{eqnarray}
\label{rep2}
S&=&{1\over\Delta}\int\Big(\omega(-\nabla^2\psi+V'(\psi))-
\omega^2\nonumber\\
&&\qquad\qquad+\sum_a\chi_a(-\nabla^2+V''(\psi))\chi_a\Big)d^d\!r
\end{eqnarray}
The integral over the $n-2$ commuting fields $\chi_a$ 
yields ${\rm det}(-\nabla^2+V'')^{-(n-2)/2}$ and so they 
may be replaced at $n=0$ by two anticommuting fields $\overline\chi$
and $\chi$.
The supersymmetry is made explicit by introducing anticommuting
coordinates $(\theta,\overline\theta)$ and a superfield
$\Psi\equiv\psi+\frac12(\overline\theta\chi+\theta\overline\chi)
-\frac14\overline\theta\theta\omega$,
whence $S$ may be written
\begin{equation}
\label{susyS}
S={1\over\Delta}\int\left(\ffrac12\Psi(-\nabla^2_{SS})\Psi
+V(\Psi)\right)d^d\!rd\theta d\overline\theta
\end{equation}
where
$\nabla^2_{SS}=\nabla^2+4\partial^2/\partial\theta\partial\overline\theta$.
This exhibits supersymmetry under rotations which leave
$r^2+\theta\overline\theta$ invariant. Parisi and Sourlas\cite{PS}
argued that this theory
exhibits a remarkable property of dimensionality
reduction (for a nonperturbative proof see Ref.~\cite{JCsusy}):
correlation functions whose arguments are restricted to a $d-2$-dimensional
subspace are the same as those for a non-supersymmetric theory in $d-2$
dimensions, whose action is 
\begin{equation}
\label{Sprime}
S_{\rm YL}={1\over\Delta}\int\left(\ffrac12\psi(-\nabla^2)\psi
+V(\psi))\right)d^{d-2}\!r
\end{equation}
where in this case $V(\psi)=t\psi-\frac13\psi^3$. There is one
subtlety: before dropping the irrelevant terms, the contour in $\psi$
should be rotated, in this case parallel to the imaginary axis, so as to
make the integral defined nonperturbatively. The potential therefore
becomes ${\rm i}t\psi+\frac13{\rm i}\psi^3$. 
Thus (\ref{Sprime})
is just the action for the scaling theory of the Yang-Lee edge
singularity\cite{LY}, as discussed by Fisher\cite{FisherLY}.
From (\ref{susyS}) it is seen 
that $\Delta$ has dimension $({\rm length})^{-2}$ and
this is not affected by loop corrections, otherwise supersymmetry would
be broken\cite{noteRF}. 
It flows to infinity under the RG, and $\Delta^{-1}$
is irrelevant. However, it is a classic example of a \em dangerously
\em irrelevant variable: it cannot be set equal to zero in
the scaling theory. It is responsible for the modified
hyperscaling relation $2-\alpha_{BP}=(d-2)\nu_{BP}$. 

Based on the above considerations, it is reasonable to conjecture that,
up to possible constants, $\Delta$ and $g$ should be identified,
as should $x_c-x$ and $t$\cite{note1}.
Thus, apart from non-universal constants, $G^{(r)}_{\rm sing}$ is
given by the one-point function $\langle\Psi\rangle$ in the
supersymmetric theory (\ref{susyS}),  which, by dimensional reduction,
is the same as the one-point function $\langle\psi\rangle$ in the
Yang-Lee scaling theory (\ref{Sprime}). For $d=2$ the gradient terms are
absent, so
$G^{(r)}_{\rm sing}(x,g)=b_1(gd/dx)\ln Z_1$
where
\begin{displaymath}
Z_1=\int_{-{\rm i}\infty}^{{\rm i}\infty}
{\rm e}^{(b_2/g)(-(x_c-x)\psi+\frac13\psi^3)}d\!\psi
\end{displaymath}
and $b_1$ and $b_2$ are non-universal constants. After rescaling the
integration variable, this gives the main results (\ref{scaling},\ref{airy}) of
RBG, together with the values $\theta=\ffrac13$, $\phi=\ffrac23$ for the
exponents. 

According to (\ref{airy}), the scaling function $F(s)$ has
singularities at the zeros of the Airy function, which lie on the
negative real axis. The closest to $x=0$ lies at
$x_c(g)=x_c+(2.388\ldots)(g/b_2)^{2/3}$, 
governing the asymptotic behavior $\sim x_c(g)^{-N}$
of $\sum_Ap^{(r)}_{N,A}{\rm e}^{-gA}$ as $N\to\infty$, for fixed small $g$. 
This singularity is a simple pole, corresponding to the value
$\alpha_{BP}\equiv3-\theta_{BP}=2$.
All this agrees with 
general crossover theory\cite{Fisher1} that the scaling
function should exhibit the critical singularities of the stable fixed
point.

We now discuss the generalization to
area-weighted two-dimensional self-avoiding loops at higher order
multicritical points.
The additional interactions between nearby portions of the
self-avoiding loop will modify the parameters $u_p$ in (\ref{replica}).
As long as the truncation of terms leading to (\ref{rep2}) remains
valid, the dimensional reduction argument still applies with a
modfied potential $V$,
so that the rooted generating function is still given by the logarithmic
derivative of a generalized Airy integral of the form 
$\int_C{\rm e}^{-V(\psi)/g}d\!\psi$.
The obvious candidates
for potentials which then yield multicritical behaviour in the
limit $g\to0$ have the form\cite{comment}
$V(\psi)=\sum_{j=1}^kv_j\psi^j-\psi^{k+2}$ (the coefficient of
$\psi^{k+1}$ is redundant, as it can be removed by a shift in $\psi$.)
Here $v_1$ is linear in $x_c-x$, and one can check that the other
coefficients $v_j$ should be positive deep inside the single phase region.
Repeating the above analysis then leads to the result
$G^{(r)}_{\rm sing}(v_j,g)=g(d/dv_1)\ln Z_k$ where
\begin{equation}
\label{Zk}
Z_k=\int_C{\rm e}^{-(\sum_{j=1}^kv_j\psi^j-\psi^{k+2})/g}d\!\psi
\end{equation}
with the contour $C$ chosen to guarantee convergence.
Comparing with the scaling form (\ref{kscaling}) then gives the results
(\ref{yjk}), together with
$\theta_k=1/(k+2)$. In particular $\langle R^2\rangle\sim\langle
N\rangle^{2\nu_k}$, where
$\nu_k=1/y_1(k)=(k+2)/2(k+1)$.
It should be noted
that, although these exponent values are based on extremizing a simple
polynomial, they are not the same as those in Landau 
theory, in which the analogous potential would be 
$\sum_{j=1}^kv_j\phi^{2j}+\phi^{2(k+1)}$.

Finally, the limit $g\to0$ may be taken in (\ref{Zk}), using the
saddle-point method, with the result that $G^{(r)}_{\rm sing}$
is simply given by 
a zero of $V'(\psi)$. By considering the limit when all the $v_j$ are
large and positive, it may be shown that the correct zero in this single
phase region is that on the real axis with the largest real part.
The contour is to be run through this, locally parallel to the imaginary
axis.
Thus, near the $k=2$ multicritical point,
where $V(\psi)=v_1\psi+v_2\psi^2-\psi^4$,
one finds the result (\ref{thetascaling}), as the appropriate root of a cubic
equation.
This formula has a number of interesting properties. At $v_2=0$, 
$G^{(r)}_{\rm sing}$ behaves like $(x_c-x)^{1/3}$. For fixed $v_2>0$,
the first singularity occurs not at $s=1$, but at $s=-1$, and this
is a square root: $G^{(r)}_{\rm sing}\sim(x_c(v_2)-x)^{1/2}$,
where $x_c(v_2)-x_c(0)\sim v_2^{3/2}$, all as expected on the basis of
crossover theory. When $v_2<0$, the corresponding scaling function is
$\Phi^{<}(s)=\big((s^2+1)^{1/2}+s\big)^{1/3}-
\big((s^2+1)^{1/2}-s\big)^{1/3}$, where now $s=v_1/(-v_2)^{3/2}$. 
The other two roots remain complex, and
there is now a first-order transition at $s\approx1.6$, when ${\rm Re}\,V$
has the same value at each root. Note that 
$\Phi^{<}$, continued into the phase coexistence region,
is analytic near the real axis and does not exhibit a spinodal singularity.

However, there is a puzzle, which our theory has in common with that
of Saleur\cite{saleur}, associated with the physical identification
of this multicritical point. The generalisation of the Flory
approximation\cite{Flory}
to the $\Theta$ and higher-order multicritical points, at
which the first $k$ renormalised virial coefficients vanish,
may be phrased as follows: consider a long loop of length
$N$ and linear size $R$. In the absence of interactions, its entropy
may be estimated on the basis of a free random walk to be $O(R^2/N)$.
The mean density is $O(N/R^2)$, so that the interaction energy may
be approximated by $v_{k+1}N(N/R^2)^k$, where $v_{k+1}$ is the first
non-vanishing virial coefficient. Balancing these two contributions to
the free energy then gives $R\sim N^{\nu_k}$ with $\nu_k$ as above.
This argument, combined with observation that at the $k$th
multicritical point there should be exactly $k$ relevant parameters,
suggests that $k=2$ in our theory should be identified with the
$\Theta-$point, and that, because of the simple scaling associated with
the dimensionally reduced theory, the Flory result is in fact exact.
However, the value $\nu_2=\frac23$, and the associated crossover
exponent $y_2(2)/y_1(2)=\frac23$, do not agree with the predictions of
an exactly solvable model of Duplantier and Saleur\cite{DS}, for which
the corresponding values are $\frac47$ and $\frac37$. 
While it might be argued that
this model is somewhat special, extensive numerical studies of more
generic lattice models appear to confirm these values\cite{debate}.
Even more strikingly, the supposedly correct value for $\nu$ at the
$\Theta$-point actually corresponds to $k=6$ in our theory (and that of
Saleur), and the crossover exponent corresponds to a perturbation with
$j=5$. But this relevant perturbation should lead to a multicritical
point with $k=3$, not the usual one with $k=1$ as expected on the
physical grounds for the $\Theta$-point. This same paradox was present
in Saleur's theory\cite{saleur} and at present there seems to be no
plausible resolution. It may well be that the coincidence of the
$\Theta$-point exponents with $k=6$ is merely that, and that the
sequence of multicritical points of self-avoiding loops implied by
Saleur's and the present theory represent some completely different
physics. Unfortunately, because of the truncations made in going from
the original self-avoiding loop model to the generalised Airy integral,
it is very difficult to say to what these other multicritical points might
correspond physically.

To summarize, the conjectured scaling function of Richard, Guttmann and
Jensen\cite{RGJ} for area-weighted
self-avoiding polygons has been shown to follow from physical reasoning
concerning the crossover to branched polymers, together with the
dimensional reduction argument of Parisi and Sourlas\cite{PS}. Depending
on ones point of view, the numerical confirmation of this formula found
by RGJ could be taken as dramatic vindication of the dimensional
reduction argument, beyond its simple prediction of the value of the
entropic exponent $\alpha_{BP}=2$. The exact formula for the scaling
function is in accordance with standard crossover theory\cite{Fisher1},
but it points to the importance of understanding all the singularities
of the crossover scaling function, not just the physical ones,
in building up the full scaling form.

The formula proposed by RJG is just the first of a series of exact
scaling functions describing higher-order multicritical points for
self-avoiding loops weighted by their area. Moreover this approach
enables one to recover exact results for scaling functions in the
unweighted ensemble, and these have the form of algebraic functions.
These are the first examples of exact but nontrivial scaling functions
of more than one thermodynamic variable at isotropic critical points.

The simple structure found here is analogous to that which appears in $N=2$
supersymmetric theories in two dimensions, although in this
case the supersymmetry is of a different nature. From that point of
view the multicritical points of self-avoiding loops 
correspond to the $A_{k+1}$ series of simple singularities\cite{catas}
of the potential $V$: it would be interesting to find analogs of the
$D_{k+1}$ series, and the exceptional cases.

The author thanks the authors of Ref.~\cite{RGJ} for sending a draft copy
of their paper prior to publication, and A.~Owczarek and H.~Saleur for
their comments on an earlier version of this manuscript.
This research was supported in part by the Engineering and
Physical Sciences Research Council under Grant GR/J78327.

\end{multicols}

\begin{references}
\bibitem{note1} The above argument implies only that $t\propto x_0-x$, 
but in fact $x_0=x_c$ for consistency with (\ref{scaling}).
\bibitem{noteRF} This is presumably what happens in the random field
Ising model, which does not obey simple dimensional reduction.
\end{references}
\end{document}